\documentclass[10pt,twocolumn]{IEEEtran}
\ifCLASSINFOpdf
\else
\fi

\usepackage{amssymb}
\usepackage{amsmath}
\usepackage{amsthm}
\usepackage{graphicx}
\usepackage{algorithm}
\usepackage{algorithmic}
\usepackage{booktabs}
\usepackage{cite}
\usepackage{bm}
\usepackage{float}
\usepackage{multirow}
\usepackage{color}
\usepackage{subfigure}
\usepackage{caption}
\usepackage{epstopdf}
\usepackage{hyperref}
\usepackage{fixltx2e}
\usepackage{longtable}
\usepackage{diagbox}
\usepackage{changepage}
\usepackage{comment}
\usepackage{cases}
\usepackage[short]{newoptidef}
\usepackage{makecell}
\usepackage{mathabx}
\theoremstyle{definition}

\theoremstyle{remark}

\theoremstyle{plain}

\begin{document}
		%
		\title{Holographic-Type Communication for Digital Twin: A Learning-based Auction Approach}
		
		
		
		%
		\author{\IEEEauthorblockN{XiuYu Zhang,
				Minrui Xu,
				Rui Tan, and
				Dusit Niyato, \emph{Fellow, IEEE}}
			
			\IEEEauthorblockA{School of Computer Science and Engineering,
				Nanyang Technological University,
				Singapore}
				
			
			}


		\maketitle
		\thispagestyle{empty} 
        \pagestyle{empty}    
		\begin{abstract}
		Digital Twin (DT) technologies, which aim to build digital replicas of physical entities, are the key to providing efficient, concurrent simulation and analysis of real-world objects. In displaying DTs, Holographic-Type Communication (HTC), which supports the transmission of holographic data such as Light Field (LF), can provide an immersive way for users to interact with Holographic DTs (HDT). However, it is challenging to effectively allocate interactive and resource-intensive HDT services among HDT users and providers. In this paper, we integrate the paradigms of HTC and DT to form a HTC for DT system, design a marketplace for HDT services where HDT users' and providers' prices are evaluated by their valuation functions, and propose an auction-based mechanism to match HDT services using a learning-based Double Dutch Auction (DDA). Specifically, we apply DDA and train an agent acting as the auctioneer to adjust the auction clock dynamically using Deep Reinforcement Learning (DRL), aiming to achieve the best market efficiency. Simulation results demonstrate that the proposed learning-based auctioneer can achieve near-optimal social welfare at halved auction information exchange cost of the baseline method.     
		\end{abstract}

		%
		\IEEEpeerreviewmaketitle

		\section{Introduction}
		

        
        Digital Twin (DT) is a digital replica of a physical entity in the physical world, which contains all or most of the information about the physical entity. Typically, each DT system consists of three components: a physical entity, a corresponding virtual/digital model of the physical entity, and a link/channel connecting the physical entity and virtual model~\cite{Fuller_Fan_Day_Barlow_2020}. In DT systems, constructing and updating the virtual model requires simultaneous retrieval and analysis of information from the physical entity~\cite{9120192,8901113}. Then actuators can make changes to the physical entity's state after the decision-making that involves the reverse direction of the data flow. Through the bi-directional communications between the virtual model and the physical entity, DT facilitates informed decision-making in many domains, e.g., data centers, factories, and smart cities. For instance, DT can help enterprises design, construct better physical data centers, and improve the energy efficiency of the data center operations~\cite{Wang_Zhou_Dong_Wen_Tan_Chen_Wang_Zeng_2020}. 
        
        Representation is one of the central issues in DTs to improve users' experience during interactions with DTs. The virtual representation of a DT can take various forms independently of the original physical entity, e.g., numerical data, images, videos, and holograms. Although technologies to capture, render, and display traditional two-dimensional (2D) images and videos are mature, they are incapable of presenting visuals immersively, like how the human visual system perceives the world, due primarily to the lack of depth information in 2D representation. Fortunately, this limitation of traditional imaging motivates holographic technologies~\cite{10.1145/3386569.3392485}. DT with holographic visual representation~\cite{hdt_2020}, i.e., Holographic Digital Twin (HDT), can provide a life-like realistic, immersive, and interactive experience to the users~\cite{hdt_classroom}. Therefore, Holographic‑Type Communication (HTC) supported by 6G and beyond will be an enabling technology for HDT to display multi-view 3D holograms recorded in the form of a Light Field (LF) with high resolution and framerate~\cite{Akyildiz_Guo}. 
        
\begin{figure}[t]
			\centering
			\includegraphics[width=1\linewidth]{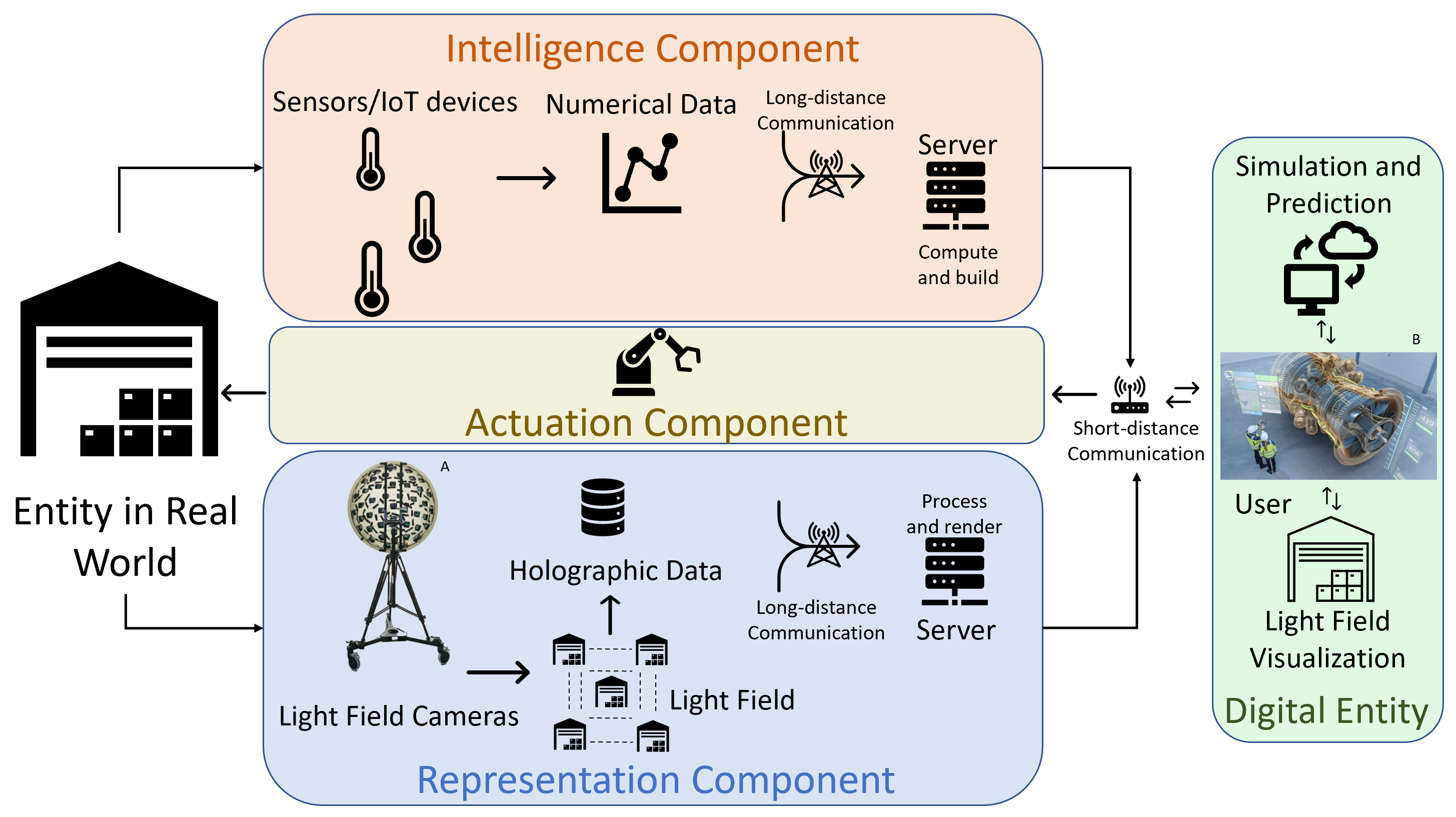}
			{\tiny{Visual A, B are adapted from~\cite{10.1145/3386569.3392485} and~\cite{diagram_b}}.}
			\caption{The Interplay of HTC and DT}
			\label{fig:HDT}
		\end{figure}

        
        Light Field, one of HTC displays introduced in~\cite{lightfield} to describe radiometric properties of light, is often represented as a vector function that defines the quantity and behavior of light rays in a three-dimensional (3D) space. In an LF, the geometrical information of a light ray is mapped to the attributes of light, such as RGB values. Thus, the depth and parallax of real-world objects are preserved~\cite{Akyildiz_Guo}. The visual representation of HDTs is captured and processed as LFs, as shown in Fig.~\ref{fig:HDT}. The physical entity is first captured, using LF cameras and related equipment, and coded into holographic data suitable for transmission~\cite{Akyildiz_Guo}. The holographic data received by the server is rendered as LF video with the requested angular resolution for users to view. The rendered holographic visuals can be displayed in various types of display, including XR head‑mounted display, multi‑view volumetric display, and LF display~\cite{Akyildiz_Guo}. 
        
        This paper considers a display-agnostic HDT system with the following three characteristics: 1) resource estimation, 2) multiple viewpoints, and 3) interactivity. A HDT’s LF virtual representation can be coded and viewed as an LF video with multiple viewpoints in every frame and each frame is effectively rendered  in 3D by collectively displaying depth information~\cite{10.1145/3386569.3392485}. Moreover, considering the duration of the video, one may estimate the minimum amount of resources required to support the display of a HDT’s virtual representation for a specific duration of time. Furthermore, multiple viewpoints allow each viewer to observe and interact with a HDT from different angles. The current workflow to create and view an LF is time-, computation-, and communication-intensive. Several researchers attempted to address each of these issues. For example, capturing an LF requires multiple plenoptic cameras, which can now be reduced by learning-based approaches to only a regular camera or fewer photos~\cite{10.1145/3072959.3073614, Inagaki_2018_ECCV}. Processing, rendering and streaming LF as an image or video need significant computing and communication resources, which can now be alleviated by new acceleration and compression algorithms~\cite{mdc, roi}. Therefore, LF is promising for forming a holographic visual representation of DTs. However, the process of capturing and processing an LF is still resource-consuming compared with the traditional imaging approaches. Therefore, in order to utilize the characteristics of an LF for representing a DT, especially when multiple users request to access/view the visual representation simultaneously with varying quality expectations, an efficient way to allocate scarce computing resources to fulfill the needs of as many users as possible is required. A simplistic allocation algorithm will result in resource wastage and even failures to deliver HDT with the support of 6G communication.
		
		In this paper, we propose a HTC for DT system, where HDTs are displayed in the form of an LF. On the one hand, we consider HDT users' Volume of Interest (VoI) in a HDT, i.e., both spatially and temporally representing the demand for HDT. On the other hand, HDT service providers need to consume both transmission and computational resources to provide real-time rendering and transmission of HDT. To achieve the equilibrium between demand and supply in a HDT's bilateral market, we propose a Double Dutch Auction (DDA)-based mechanism for matching and pricing HDT users and providers. Finally, to improve the efficiency of the proposed outcry auction, we adopt a Deep Reinforcement Learning (DRL)-based mechanism for the auctioneer to adjust auction clocks dynamically for users and providers. Specifically, the auction process is formulated as a Markov Decision Process (MDP), in which the learning agent acts as the auctioneer to learn efficient action clock adjustment during interaction with the HDT market environment.
		
		Our contributions can be summarized as follows.
		\begin{itemize}
		    \item This is the first paper that integrates the paradigms of holographic-type communication and digital twin such that the visual representation of digital twin is a light field. Holographic digital twins with light field representation provide a better perceptual immersive experience and thus improves the access and utilization of digital twin.
		    \item We design the marketplace for holographic digital twin services, where the valuation function of users considers time-depending preferences with objective opinion scores, and the valuation function of providers consider both communication and computation costs.
		    \item We propose an intelligent mechanism where an auctioneer is trained using deep reinforcement learning for the holographic digital twin market based on the double Dutch auction, which effectively reduces the communication cost while preserving social welfare.
		\end{itemize}
		
		\section{Market Design and System Model}
		
		\subsection{The HDT Market}
		We consider a HDT system with a set $\mathcal{M} = \{1, \ldots, m, \ldots, M\}$ of $M$ HDT users and a set $\mathcal{N}=\{1,\ldots, n, \ldots, N\}$ of $N$ HDT providers. The system aims to provide HDT services of monitoring the physical entity from HDT providers to users. As every user may want to view the virtual representation of the HDT for different purposes, their valuation of the service received might be different. In the meanwhile, every provider might have different hardware and cloud resources, so their valuation of the service provided can be different. We consider this matching and pricing problem in a market, where the buyers are the users, and the sellers are the providers. 
		
		\subsubsection{HDT Visual Representation Viewer (Buyer)}
		The buyers in the market pay the sellers to access the visual representation. The viewers submit their buy-bid to the auctioneer before the start of every round in the auction. The buy-bid of viewer $m$ at round $t$ is evaluated as $k^t_m = V_m^B(R_m, \mathbf{a}_m, d_m, j_m)$, i.e., the maximum price that buyer $m$ is willing to pay for the visual representation. The viewer $m$ can specify the quality of the requested visual representation by providing the resolution $R_m$, interest in each viewpoint $\mathbf{a}_m$, and the duration of viewing $d_m$. The valuation of the buyer $m$ can also be affected by their interest decay rate $j_m$. 
		
		\subsubsection{HDT Visual Representation Provider (Seller)}
		The sellers in the market are HDT providers that capture and deliver the visual representation to the viewers. The providers submit their sell bid to the auctioneer before the start of every round in the auction. The sell-bid of provider $n$ is evaluated as $l^t_n = V_n^S(r_n, e_n, c_n, f_n, a_n, d_n)$, i.e., the minimum price that seller $n$ is willing to accept for providing the service. The provider $n$ can specify its ability using its base rate $r_n$, spectrum efficiency $e_n$, CPU cycles to execute $c_n$, CPU frequency $f_n$, default angular resolution $a_n$, and default service duration $d_n$. 
		
		\begin{table*}[!]
\small\centering
\caption{The objective opinion score under different transmission rates supports different HDT resolutions and framerates of a viewpoint.}
\begin{tabular}{|c|c|c|c|c|c|}
\hline
Bitrates/Mbps & $\geq$ 21 & $\geq$ 55 & $\geq$ 125 & $\geq$ 221 & $\geq$ 529 \\ \hline
Resolution              & 720$\times$480 & 1280$\times$720 & 1920$\times$1080 & 2560$\times$1440 & 4080$\times$2160 \\ \hline
FPS                     & 60      & 60       & 60        & 60        & 30        \\ \hline
Objective opinion score & 1       & 2        & 3         & 4         & 5         \\ \hline
\end{tabular}
\label{tab:objective}
\end{table*}
		\subsection{Valuation Model}

		\subsubsection{Valuation of Buyers}
\begin{figure}
			\centering
			\includegraphics[width=0.9\linewidth]{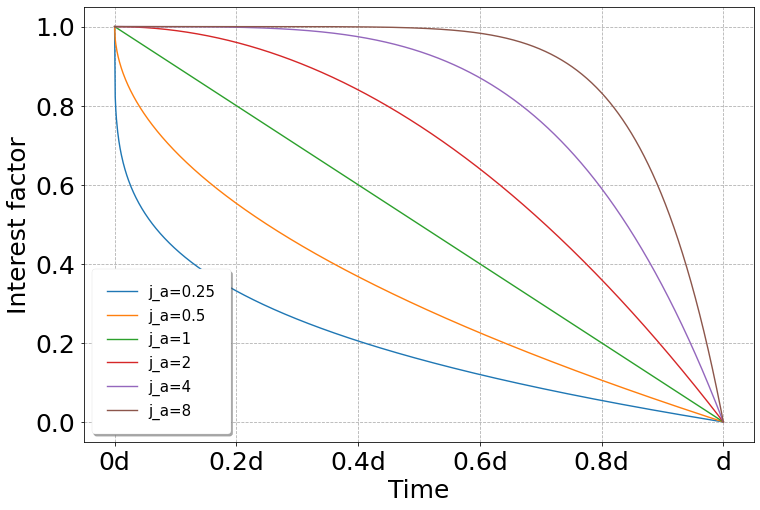}
			\caption{Effect of $j_a$ on the Interest Factor}
			\label{fig:interest_factor}
		\end{figure}

		A buyer in the market is a user of the HDT who requests to view the visual representation of the HDT. The demand value of buyer $m$ is expressed as
		\begin{equation}
		    V_m^B = g_m^B\big(o_m(R_m) \times \emph{VoI}_m(\mathbf{a}_m, d_m, j_m)\big),
		\end{equation}
		where $g_m^B(\cdot)$ is an increasing and concave function, $\emph{VoI}_m(\cdot)$ is the VoI of buyer $m$, and $o_m(\cdot)$ is the objective opinion score of buyer $m$ as defined in TABLE~\ref{tab:objective}. The VoI of a viewer can be computed as 
		\begin{equation}
		    \emph{VoI}(\mathbf{a},d, j_a) =\sum_{i=1}^{|\mathbf{a}|}\sum_{t=1}^d (1 - (\frac{t}{d})^{j_a})\times \mathbf{a}_i
		\end{equation}
		where $\mathbf{a}$ is a vector representing the viewer's interest in each viewpoint of the visual representation, $d$ is the intended viewing duration, and $j_a$ is the viewer's interest decay factor. $\emph{VoI}$ measures the total amount of interest of the viewer throughout viewing and across all viewpoints. The $j_a>0$ reflects the decay of the interest factor $1 - (\frac{t}{d})^{j_a}$, i.e., the smaller the $j_a$, the faster the rate of decrease, as shown in Fig.~\ref{fig:interest_factor}.
		
		The computed demand value needs to go through a transformation, i.e., $g_m^B(\cdot)$, which maps the unbounded value to the designed auction price range to be used as the buy-bid. Based on the rule of DDA, the buyer with a higher bid is first considered and possibly matched. 
		
		
		\subsubsection{Valuation of Sellers}
		A seller in the market is a provider that provides a visual representation of the HDT. In the process HDT tasks, per data size of HDT tasks requires $c_n$ CPU cycles to execute. Let $e_n$ denote the spectrum efficiency of seller $n$ and $f_n$ denote the CPU frequency of seller $n$. The supplied value of seller $n$ after conversion is computed as
		\begin{equation}
		    V_n^S=g_n^{S,com}\bigg(\frac{r_n}{e_n} a_n d_n\bigg) + g_n^{S,cmp}\bigg(\frac{r_nc_n}{f_n} a_n d_n\bigg),
		\end{equation}
		where $g_n^{S,com}(\cdot)$ and $g_n^{S,cmp}(\cdot)$ are decreasing and convex functions, $r_n$ is the base rate of this seller, $a_n$ is the default number of interactive points (angular resolution) provided by the seller, and $d_n$ is the default service duration provided by the seller. The supplied value after transformation, i.e., $g_n^S(\cdot)$, which maps the unbounded value to the designed auction price range, is used as the sell-bid. Based on the rule of DDA, the seller with a lower bid is first considered and possibly matched. Collectively with the buyer's bid design, the buyers with a higher demand value are matched to the sellers with a higher supply value. 
		


		\subsection{Double Dutch Auction}
		By applying the DDA mechanism to a HDT market, we aim to maximize the collective utility of buyers and sellers. The auctioneer in the market is an algorithm integrated into the virtual representation of the HDT that initializes and adjusts two Dutch clocks $C_B, C_S$, and one auction flag $\Psi$ used in the DDA. 
		The buyer clock $C_B$ shows the current buying price in each round in the auction, and it starts with the highest price allowed in the auction $p^{max}$ and descends every buyer round. On the other hand, the seller clock $C_S$ shows the current selling price in each round in the auction,  and it starts with the lowest price allowed in the auction $p^{min}$ and increments every seller round. Furthermore, the auction clock determines the owner of any auction round $t$: buyer round and seller round if the auction flag $\Psi^t$ is $0$ and $1$, respectively. 
		
		The auction starts with the buyer round, i.e., $\Psi^0=0$. When the auction starts, the $M$ buyers are arranged in non-increasing order, i.e., $B=\{m|0\leq m < M \land m<k \implies v_m^B \geq v_k^B\}$, and the $N$ sellers are arranged in non-decreasing order, i.e., $S=\{n|0\leq n < N \land n<k \implies v_n^S \leq v_k^S\}$, based on their valuation. All the buyers and sellers are active when the auction starts, i.e., they have yet to accept any Dutch clock price that $B^0_{active}=B$ and  $S^0_{active}=S$. This also means that the sequences of winning buyers $B_{winning}^0$ and sellers $S_{winning}^0$ are empty when the auction starts. Throughout the auction at every round $t$, $B = B^t_{active} \cup B^t_{winning}, B^t_{active} \cap B^t_{winning} = \emptyset$ and $S = S^t_{active} \cup S^t_{winning}, S^t_{active} \cap S^t_{winning} = \emptyset$. In addition to these sequences, two sequences $O_B$ and $O_S$ are used to record the accepted prices of two Dutch clocks, and sequence $N_A(t)$ is used to record the number of audiences of the broadcast at every auction round $t$. During an auction round $t$, market participants decide whether to accept the clock price while the auctioneer performs four actions. 
        \subsubsection{Auctioneer Broadcasts Dutch Clock} 
        The auctioneer broadcasts the buyer clock value $C^t_B$ to the active buyers if the auction flag $\Psi^t=0$, and the seller clock value to active sellers if the auction flag $\Psi^t=1$.
        
        \subsubsection{Market Participants Check Dutch Clock} 
        In the buyer's round, i.e., $\Psi^t=0$, active buyers check the received buyer clock price against their bid. If the buyer clock price value is no larger than a buyer's bid, that buyer $m$ accepts the buyer clock price as his/her bid, i.e., 
        \begin{equation}
            C_B^t \leq V^B_m \implies k^t_m \leftarrow C_B^t.
        \end{equation}
        The difference between the expected buy bid, i.e.,  $k_m^t$ before the update, and the actual bid, i.e., $k_m^t$ after the update, is the regret of buyer $m$.
        On the other hand, in the buyer's round, i.e., $\Psi^t=1$, active sellers check the received seller clock price against their bid. If the seller clock price value is no smaller than a seller's bid, that seller $n$ accepts the seller clock price as their bid, i.e.,
        \begin{equation}
            C_S^t \geq V^S_n \implies l^t_n \leftarrow C_S^t.
        \end{equation}
        The difference between the expected buy bid, i.e.,  $l_n^t$ before the update, and the actual bid, i.e., $l_n^t$ after the update, is the regret of seller $n$.
        Note that multiple market participants can accept the same clock price in the same round. The validity of acceptance is checked when the auction ends. The number of audiences is added to the sequence $N_A$.
        
        \subsubsection{Auctioneer Records the Acceptance of Dutch Clock}
        In the buyer's round, i.e., $\Psi^t=0$, all buyers who accepted the current buyer clock price are marked inactive, i.e., for all those buyers $m$, 
        \begin{equation}
            B_{active}^{t+1} = B_{active}^{t} \setminus \{m\},
        \end{equation}
        and added to the sequence of winning buyers, i.e.,
        \begin{equation}
            B_{winning}^{t+1} = B_{winning}^{t} \cup \{m\}.
        \end{equation}
        The accepted price of the buyer clock will also be added to $O_B$. Then the flag is changed, i.e., $\Psi^{t+1}=1$ if there are active sellers. Similarly, in the seller's round, all the sellers who accepted the current buyer clock price are marked inactive, i.e., for all those sellers $n$,
        \begin{equation}
            S_{active}^{t+1} = S_{active}^{t} \setminus \{n\},
        \end{equation}
        and added to the sequence of winning sellers, i.e.,
        \begin{equation}
            S_{winning}^{t+1} = S_{winning}^{t} \cup \{n\}.
        \end{equation}
        The accepted price of the buyer clock will also be added to $O_S$. If multiple market participants accept the same clock price at the same round, then that clock price will be added to the corresponding sequence multiple times. Then the flag is changed, i.e., $\Psi^{t+1}=0$ if there are active buyers.
        
        \subsubsection{Auctioneer Adjusts Dutch Clock}
        The Dutch clock is adjusted by a chosen stepsize which is a multiple of the minimum price interval $p^*$, i.e., $\Theta^t=kp^*$ where $k$ is a positive integer. The auctioneer descends the price of the buyer clock in the buyer's round by the stepsize, i.e., 
        \begin{equation}
            \Psi^t=0 \implies C_B^{t+1} = C_B^{t} - \Theta^t.
        \end{equation}
        On the other hand, the auctioneer increments the price of the seller clock by the stepsize in the seller's round, i.e.,
        \begin{equation}
            \Psi^t=1 \implies C_S^{t+1} = C_S^{t} + \Theta^t.
        \end{equation}
        
        \subsubsection{Auctioneer Checks the Termination Condition of the Auction}
        The auctioneer terminates the auction at round $t$ if any of the two conditions are met and records the total number of rounds in this auction as $T=t$: \begin{itemize}
            \item There are neither active buyers nor active sellers in the market, i.e., $B^{t+1}_{active}=\emptyset \land S^{t+1}_{active}=\emptyset$.
            \item The two Dutch clocks intersect, i.e., $C^{t+1}_B < C^{t+1}_S$.
        \end{itemize}
        
        After the termination of the auction, the winner determination rule is used to decide the valid pairs of deals. Let $w=\min\{|W_B|,|W_S|\}$ be the number of available candidate pairs. If the market is clear when $\Phi^T=1$, the first $w-1$ buyers and sellers win the auction, and the clearing price is set to $p^c = \frac{C_B^{T}+C_S^{T}}{2}$. If the market is clear when $\Phi^T=0$, the first $w$ buyers and sellers win the auction, and the clearing price is set to $p^c = \frac{C_B^{T+1}+C_S^{T+1}}{2}$.
		

		\subsection{Market Efficiency Metrics}
		The metric used to evaluate market efficiency is social welfare (SW), which incorporates the buyer utility, seller utility, and the resources used by the auctioneer in the auction process. As such, 
		\begin{equation}
		    SW = U_B + U_S - P_C,
		\end{equation}
		where $U_B$ is the total buyer utility, $U_S$ is the total seller utility, and $P_C$ is the total cost used in the auction, i.e., total broadcast cost. The utility of a buyer $m$ is calculated as the expense saved in accepting the price of the buyer clock as the bid instead of using his/her valuation, i.e., $p^c-O_B(m)$. The utility of a seller $n$ is calculated as the profit gained in accepting the price of the seller clock as a bid instead of using his/her valuation, i.e., $O_S(n) - p^c$. The broadcast cost in an auction round $t$ is calculated as $c_{b} \cdot N_A(t)$ where $c_{b}$ is the unit delivery cost of the broadcast. Let $w$ be the number of valid winning pairs as defined above, then
		$U_B = \sum_{i=0}^w v^B_i - O_B(i)$, $U_S = \sum_{i=0}^w O_S(i) - v^S_i$, and $P_C = \sum_{t=0}^T p_{b} \cdot N_A(t)$.
		\begin{figure*}[t]
			\centering
			\subfigure[Regret]{
			\includegraphics[width=0.3\textwidth]{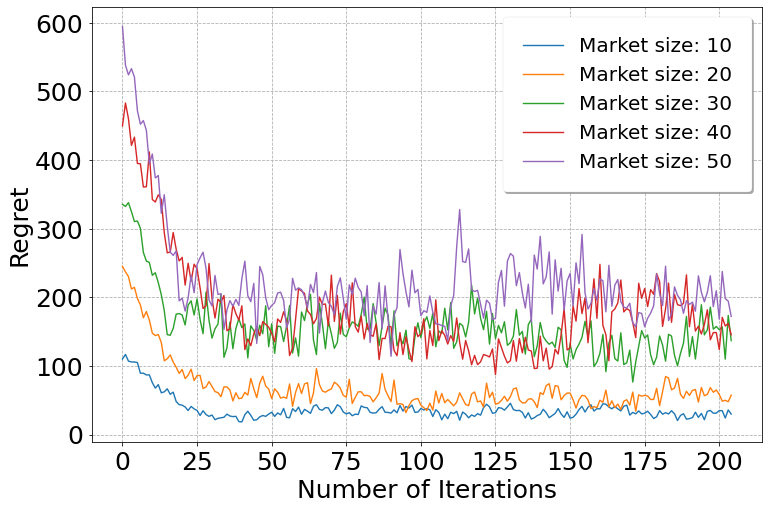}
			\label{fig:drl_regret}
			}
			\subfigure[Social welfare]{
			\includegraphics[width=0.3\textwidth]{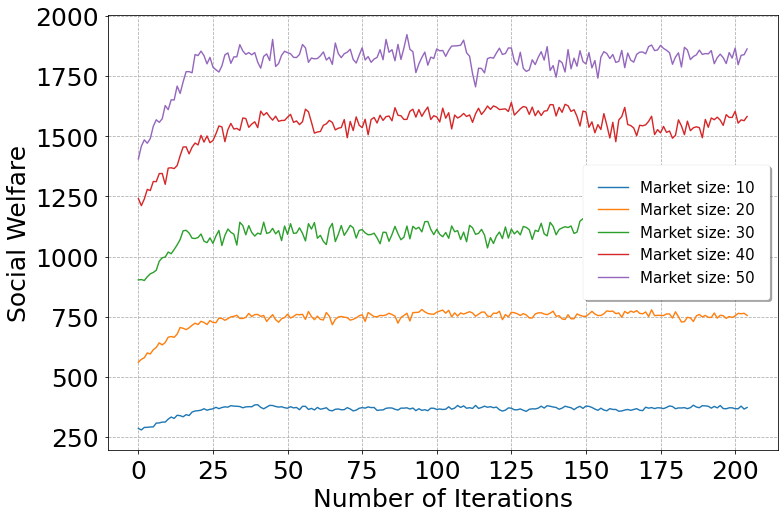}
			\label{fig:drl_sw}
			}
			\subfigure[Cost]{
			\includegraphics[width=0.3\textwidth]{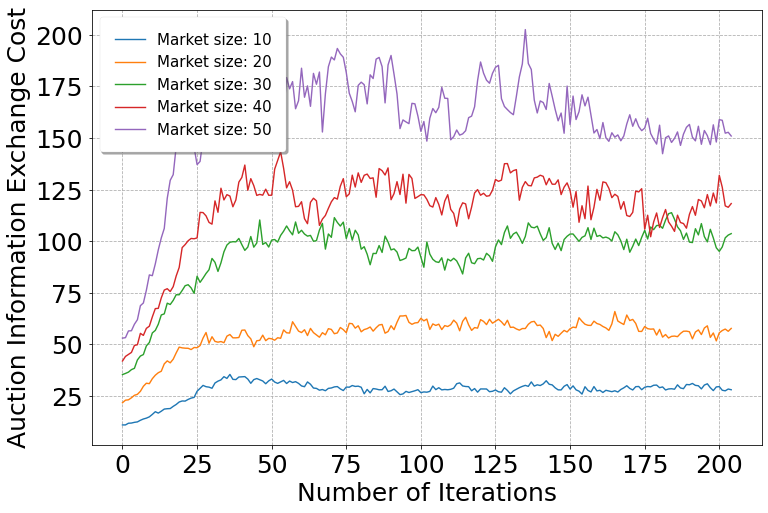}
			\label{fig:drl_cost}
			}
			
		\caption{Convergence of DRL-DDA in Different Market Sizes}
		\label{fig:drl}
		\end{figure*}
		\subsection{Problem Formulation}
		The learning-based DDA mechanism has these promising properties as follows~\cite{Xu_Niyato_Kang_Xiong_Miao_Kim_2021}.
		\begin{itemize}
            \item Individual Rationality (IR): In a HDT market, all the buyers and sellers can achieve non-negative utilities. For the losing buyers and sellers, their utilities are zero. Meanwhile, the clearing price for the winning buyers and sellers is lower than or equal to the clocks they accepted. Therefore, the learning-based DDA mechanism is IR.
		    \item Truthfulness: The auction is truthful in that all the buyers and sellers do not have the incentive to submit their buy-bids or sell-bids except for their true valuation. In the learning-based DDA, there exists a critical and monotonic payment for all the winning buyers and sellers, i.e., the common crossing price.
		    \item Budget Balance: The market is cleared with the common crossing price, which is the same for buyers and sellers. Therefore, the budget in the market is always zeros, and thus the auction is budget balance.
		\end{itemize}
		
	\section{The Intelligent Auction-based Mechanism Design}
	Reinforcement learning is used to train the agent acting as the auctioneer in the DDA with the given environment $<\mathcal{S}, \mathcal{A}, \mathcal{P}, \mathcal{R}>$ where $\mathcal{S}$ is the state space, $\mathcal{A}$ is the action space, $\mathcal{P}$ is state transition probability, and $\mathcal{R}$ is the reward.
	\subsection{Markov Decision Process for DDA}
	\subsubsection{State Space $\mathcal{S}$}
	The state space at each auction round $t$ contains the auction flag $\Psi^t$, auction round $t$, two Dutch clocks $C^t_B, C^t_S$ and the number of winning sellers and buyers, i.e., $\mathcal{S}^t = \{\Psi^t, t, C^t_B, C^t_S, |W_B^t|, |W_S^t|\}$.
	\subsubsection{Action Space $\mathcal{A}$}
	The auction space contains the possible stepsize the auctioneer can choose to adjust the two Dutch clocks, i.e., $\forall t, a^t=\Theta^t \in \mathcal{A}$. 
	\subsubsection{Reward $\mathcal{R}$}
	\begin{equation}
        r(\mathcal{S}^t, a^t, \mathcal{S}^{t+1})=
        \begin{cases}
            -u_B^t + tk_pp_C^t, &\Psi^t=0 \\
            -u_S^t +tk_pp_C^t, &\Psi^t=1
        \end{cases}
    \end{equation}
    where 
    $u_B^t=\sum_{i\in W_B^{t+1} \setminus W_B^{t}} v^B_i - C_B^t$
    is the gain in buyer utility, 
    $u_S^t=\sum_{i\in W_S^{t+1} \setminus W_S^{t}} C_B^t - v^S_i$ 
    is the gain in seller utility,
    $p_C^t=p_{b} * N_A(t)$ 
    is the information exchange cost for broadcasting, and $p_k$ is the broadcasting penalty factor.
    \subsubsection{Value Function}
	Given policy $\pi$, the value function $V_\pi(S)$ of the state $S$, the expected return when starting in $S$ and following $\pi$ thereafter, can be formally by
	$
	V_\pi(S) := \mathbb{E}_\pi\left[\sum_{k=0}^{K}\gamma^k R(S^k,a^k)|S^0=S\right]$,
	where $\mathbb{E}_\pi(\cdot)$ denotes the expected value of a random variable given that the agent follows policy $\pi$ and $\gamma\in[0,1]$ is the reward discount factor used to reduce the weights as the time step increases. Finally, to maximize the value function, the clock adjustment policy of the auctioneer is trained and evaluated with the proximal policy optimization algorithm~\cite{schulman2017proximal} via stochastic gradient ascent.
    \begin{figure*}[t]
			\centering
			\subfigure[Social Welfare]{
			\includegraphics[width=0.468\linewidth]{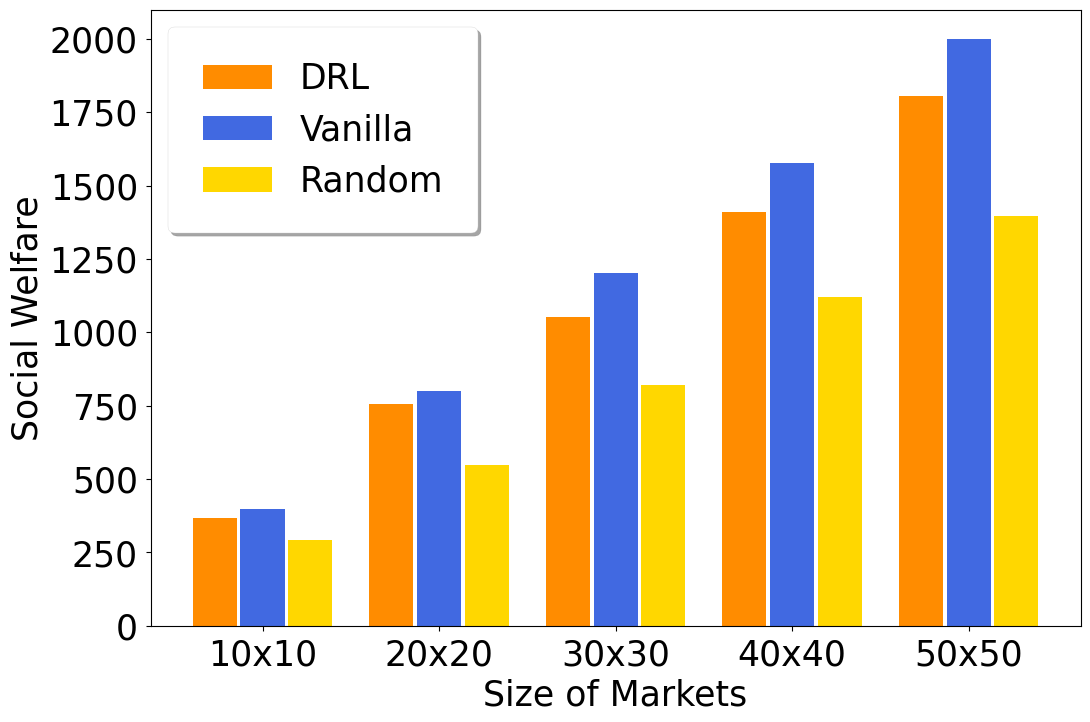}
			\label{fig:social_welfare}
			}
			\subfigure[Cost]{
			\includegraphics[width=0.465\linewidth]{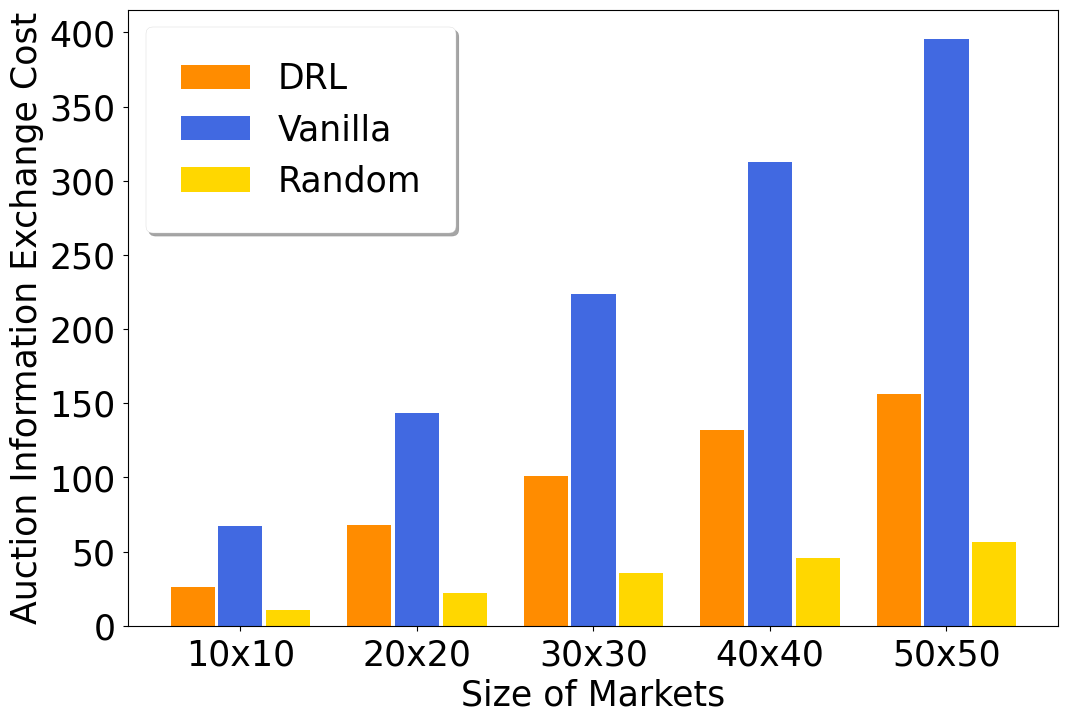}
			\label{fig:cost}
			}
		\caption{Performance of Different Auction Methods in Different Market Sizes}
		\label{fig:main_compare}
		\end{figure*}
		
	\section{Experiment}
		
	\subsection{Market Simulation}
    HDTs consisting of 10, 20, 30, 40, and 50 providers in their representation component are considered for testing the performance of the proposed learning-based auction algorithm.  In each set of simulations, an equal number of providers and viewers is assumed. This equal number of sellers and buyers is the market size in the legend of Fig.~\ref{fig:drl} and in the label of Fig.~\ref{fig:main_compare}. The markets simulated have the same number of sellers and buyers as this market design can best evaluate the performance of the auctioneers. By the winner determination rule used in the auction, the first $w$ candidate pairs will be matched and traded where $w$ is smaller or equal to the number of sellers or buyers for which the number is smaller than another. Thus the additional buyers or sellers will not contribute to the calculation of social welfare. Furthermore, the auctioneer terminates the auction when there are no active buyers or sellers, meaning that an auction in a market with an equal number of sellers and buyers can potentially take more rounds. Thus the effectiveness of the learning-based auctioneer in increasing social welfare and decreasing cost can be better tested in a market with an equal number of buyers and sellers.

    The buyers' interest vector $\mathbf{a}$ is sampled from a multivariate half-normal distribution centered at $\mathbf{1}$ with a variance of 4.  Their demand on the quality of the visual representation is uniformly sampled from the domain of the objective opinion score, as shown in TABLE~\ref{tab:objective}.  The duration of the requested access to the visual is set to between 3 minutes and 30 minutes.  The interest decay rate $j_a$ is sampled from a half-normal distribution centered at 1 with a variance of 4.  The providers' base rates, CPU cycles, CPU frequency, and spectrum efficiency are uniformly sampled from their defined domain class of $\{1,2,3\}$.  The default supply angular resolution is set to 16, and the duration is set to 15 minutes. The models used in this simulation are trained with a learning rate of $0.001$ for both actor and critic networks, and a discount factor $\gamma$ of 0.5. The policy is updated in every iteration, i.e., 2048 iterations, for 10 epochs. 
    
    \subsection{Auction Methods}
    Three auction methods are used in the experiment: Vanilla-DDA, Random-DDA, and DRL-DDA.  In Vanilla-DDA, the auctioneer adjusts the auction clocks by the minimum price interval.  In Random-DDA, the auctioneer adjusts the auctioneer clock by randomly selecting between 1 to 20 minimum price intervals.  In DRL-DDA, the auctioneer adjusts the auction clocks according to the actions proposed by the trained DRL model. As shown in Fig.~\ref{fig:drl}, when the number of iterations of the training increases, the DRL-DDA auctioneer achieves larger social welfare at the cost of increasing auction information exchange cost. This is because, as how the reward function is defined, the model is trained to minimize the regret with the applied auction information exchange penalties. The social welfare increases as the regret decrease, suggesting more information exchanges are performed with a larger cost associated. As shown in Fig.~\ref{fig:drl_regret}, the model can converge at different market sizes but with varying convergence speeds.  The larger the market size, the slower the convergence of the model.  This makes sense as when the market size is large, there are more participants in the market, and the auctioneer needs to make more decisions in the auction process.
	
	\subsection{Performance Evaluation}
	As shown in Fig.~\ref{fig:social_welfare}, the DRL-DDA outperforms the Random-DDA in all the market sizes but achieves less social welfare than the Vanilla approach.  Since Vanilla-DDA is a zero-regret method, its achieved social welfare is at the maximum.  Thus, the DRL-DDA obtains a $10\%$ lesser social welfare compared to the maximum value.  In Fig.~\ref{fig:cost}, the comparison of auction information exchange costs used by the auction methods is shown.  The Random-DDA has the least cost, i.e., only a fraction of the other methods, as it tends to finish the auction very quickly by adjusting the clocks aggressively.  The DRL-DDA uses about half of the cost of the Vanilla-DDA, suggesting that it frequently chooses to adjust auction clocks by a value larger than the one minimum price interval used by Vanilla-DDA. The performance of the DRL-DDA is expected, as we purposely trade social welfare for a smaller auction information exchange cost and a faster auction
process.
				
		
	\section{Conclusion}
	This paper presents a paradigm of HTC for DT system, which leverages LF technologies to display the visual representation of a physical entity. Specifically, LF technologies allow users to view the DT immersively and interactively, improving the usability and accessibility of the conventional DT. 
    To allocate the scarce and competing computing and communication resources required by a HDT, we formulated the problem as an economic market efficiency problem and applied DDA to it. HDT users are the buyers in the market, and the providers are the sellers in the market. Their bids used in the auction are based on the valuation of their specific demand and supply of services.
    Finally, we proposed a learning-based algorithm to act as the auctioneer in the auction, which achieves about $90\%$ of the theoretical optimum social welfare at halved information exchange cost in various market settings. 
		
		
		
		%
		\bibliographystyle{IEEEtran}
		\bibliography{ref}

	\end{document}